# The structural, elastic, electronic and optical properties of MgCu under pressure: A first-principles study

Md. Afjalur Rahman[1], Md. Zahidur Rahaman[2], Md. Atikur Rahman[*3]

[1, 2, 3]Department of Physics, Pabna University of Science and Technology, Pabna-6600, Bangladesh

## Abstract

The effect of pressure on the structural, elastic and electronic properties of the intermetallic compound MgCu with a CsCl-type structure have been investigated using *ab-initio* technique. The optical properties have been studied under normal pressure. We have carried out the plane-wave pseudopotential approach within the framework of the first-principle density functional theory (DFT) implemented within the CASTEP code. The calculated structural parameters show a good agreement with the experimental and other theoretical results. The most important elastic properties including the bulk modulus $B$, shear modulus $G$, Young's modulus $E$ and Poisson's ratio $v$ of the cubic type structure MgCu are determined under pressure by using the Voigt-Reuss-Hill (VRH) averaging scheme. The results show that the MgCu intermetallic becomes unstable under pressure more than 15 GPa. The study of Cauchy pressure and Pugh's ratio exhibit brittle nature of MgCu at ambient condition and the compound is transformed into ductile nature with the increasing of pressure. For the first time we have investigated the electronic and optical properties of MgCu. The electronic band structure reveals metallic conductivity and the major contribution comes from Cu-3d states. Reflectivity spectrum shows that the reflectivity is high in the ultraviolet region up to 72 eV.

## 1. Introduction

Magnesium (Mg) has been become an important engineering material because of its abundance in the Earth. Due to the low density (Approximately 1.74 g/cm$^3$), high specific strength, good stiffness than many other engineering materials including aluminum, steel and polymer-based composites, magnesium alloys have been attracted huge attention and gained increasing applications in transportation fields including automotive industry and aerospace manufacturing [1]. Due to these attractive properties magnesium is known as the "green" engineering material [2, 3]. Magnesium also belongs to many other attractive properties, such as electromagnetic shielding, high damping capacity, thermal conductivity, and good machinability and high recycling potential [4]. These attractive features of Mg motivated us to study this alloy. In order to improve the comprehensive properties of Mg-alloys, many theoretical and experimental attempts have been carried out and many significant progresses have been gained in recent decades. Many novel Mg-based intermetallic compounds have been prepared including MgCu with a CsCl-type structure, $Mg_{51}Cu_{20}$ with an $Mg_{51}Zn_{20}$-type structure, and $Mg_4Ni$ with an FCC structure have been reported in recent years [5]. Among them, the MgCu phase attracted huge attention since the chemical composition of MgCu is the middle of $Mg_2Cu$ and $MgCu_2$. MgCu with a CsCl-type structure have been synthesized by H. Takamura et all [6].

---

[*]Corresponding Author: atik0707phy@gmail.com

The Mg-Cu system has been studied both theoretically and experimentally [1, 6-9]. The elastic, mechanical and thermodynamic properties of MgCu intermetallic have been investigated by using density functional study [10], which shows that MgCu compound exhibits anisotropic elasticity and brittle nature at ambient pressure.

However, all the researches mentioned above are conducted at zero pressure. To the best of our knowledge the electronic and optical properties of MgCu have not been reported yet. Furthermore, the effects of pressure on the structural, elastic, and electronic properties of MgCu intermetallic compound have not been studied up to now. It is well known that pressure plays a significant role in physical properties of materials. Hence the deformation behavior of materials under compression has become quite interesting as it can provide deep insight into the nature of solid-state theories and evaluate the values of fundamental parameters [11]. It is also reported that high pressure contributes to the phase transition and changes in physical and chemical properties of material [12-14]. So the study of pressure effects on MgCu is necessary and significant. Thus, the structural, elastic and electronic properties of cubic crystal MgCu in the hydrostatic pressure range from 0 to 15 GPa and optical properties under zero pressure are investigated by the plane-wave pseudopotential density functional theory method (DFT) in the present work with the aim of having a profound comprehension about these properties. The remaining parts of this paper are organized as follows. In Section 2, the computation details are given. The results and discussion are presented in section 3. Finally, a summary of our results is shown in section 4.

## 2. Computational method

The calculations have been performed using the density functional theory (DFT) based CASTEP computer program together with the generalized gradient approximation (GGA) with the PBE exchange-correlation functional [15-19]. Mg-$2p^6 3s^2$ and Cu-$3d^{10} 4s^1$ are taken as valence electrons. The electromagnetic wave functions are expanded in a plane wave basis set with an energy cut-off of 400 eV. The k-point sampling of the Brillouin zone is constructed using Monkhorst-Pack scheme [20] with 8×8×8 grids in primitive cells of MgCu. The equilibrium crystal structures are obtained via geometry optimization in the Broyden-Fletcher-Goldfarb-Shanno (BFGS) minimization scheme [21]. In the geometry optimization, criteria of convergence are set to $1.0 \times 10^{-5}$ eV/atom for energy, 0.03 eV/Å for force, $1 \times 10^{-3}$ Å for ionic displacement, and 0.05 GPa for stress. These parameters are carefully tasted and sufficient to lead to a well converged total energy.

The elastic stiffness constants of cubic MgCu are obtained by the stress-strain method [22] at the optimized structure under the condition of each pressure. In this case, we have used generalized gradient approximation (GGA) with the PBESOL exchange-correlation function. Then the bulk modulus is obtained from the elastic constants. The criteria of convergence are set to $2.0 \times 10^{-6}$ eV/atom for energy, 0.006 eV/Å for maximum ionic force and $2.0 \times 10^{-4}$ Å for maximum ionic displacement. The maximum strain amplitude is set to be 0.003 in the present calculation.

## 3. Results and discussions

### *3.1. Structural properties*

The intermetallic MgCu compound belongs to cubic lattice of CsCl-type structure with the space group Pm-3m (221). The equilibrium lattice parameter has a value of 3.161 Å [6]. The lattice constants and atomic positions have been optimized as a function of normal stress by minimizing the total energy. The optimized structure is shown in Fig.1. The calculated values of the structural properties of MgCu are presented in Table 1 along with the available experimental and other theoretical values. From Table 1, we see that our present theoretical results are very close to both experimental and other theoretical results.

**Table 1.** The calculated equilibrium Lattice constant "$a_0$", unit cell volume "$V_0$" bulk modulus "$B_0$" and its first pressure derivative "$B_0'$" of MgCu.

| Properties | Expt.[6] | Other Calculation[10] | Present Calculation | Deviation from Expt. (%) |
| --- | --- | --- | --- | --- |
| $a_0$ (Å) | 3.161 | 3.163 | 3.21 | 1.52 |
| $V_0$ (Å$^3$) | - | 31.55 | 33.09 | - |
| $B_0$ (GPa) | - | 69.52 | 69.10 | - |
| $B_0'$ | - | 4.13 | 4.59 | - |

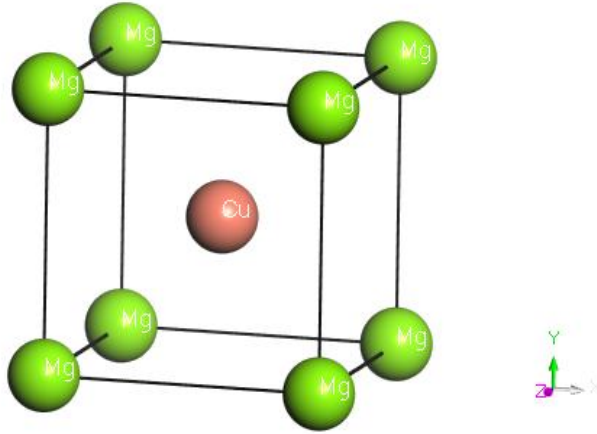

**Fig. 1.** The crystal structure of MgCu

The calculated lattice constant of this present work is 3.21 Å which shows 1.52 % deviation compared with the experimental value and slightly different than other theoretical values due to the different calculation methods. The existing discrepancy can be attributed to that our calculated data are simulated at 0 K, while the experimental data are measured at room temperature. This indicates the reliability of our present DFT based first-principles calculations. To study the influence of external pressure on the crystal structure of MgCu, we have studied the variations of the lattice parameters and unit cell volume of MgCu with different pressure up to 15 GPa. In Fig. 2(a), the variations of the cell volume and lattice parameters of MgCu with pressure are presented. It can be noticed that the ratio $a/a_0$ and normalized volume $V/V_0$ decreases with the increase of pressure, where $a_0$ and $V_0$ are the equilibrium lattice parameter and volume at zero pressure respectively. However, with the increase of pressure, the distance between atoms is reduced. As a result the repulsive interaction between atoms is strengthened, which leads to the difficulty of compression of the crystal under pressure. In order to calculate the bulk modulus $B_0$ and its pressure derivative $B_0'$ a complete geometry optimization for the lattice cell in the pressure range from 0 GPa to 15 GPa with a step of 5 GPa is performed. The obtained pressure-volume data are fitted to a third-order Birch-Murnaghan equation of state (EOS) [23]. The pressure-volume curves of MgCu intermetallic from the calculation are plotted in Fig. 2(b).

The Bulk modulus $B_0$ and its pressure derivative $B_0'$ is listed in Table 1 with available other theoretical data.

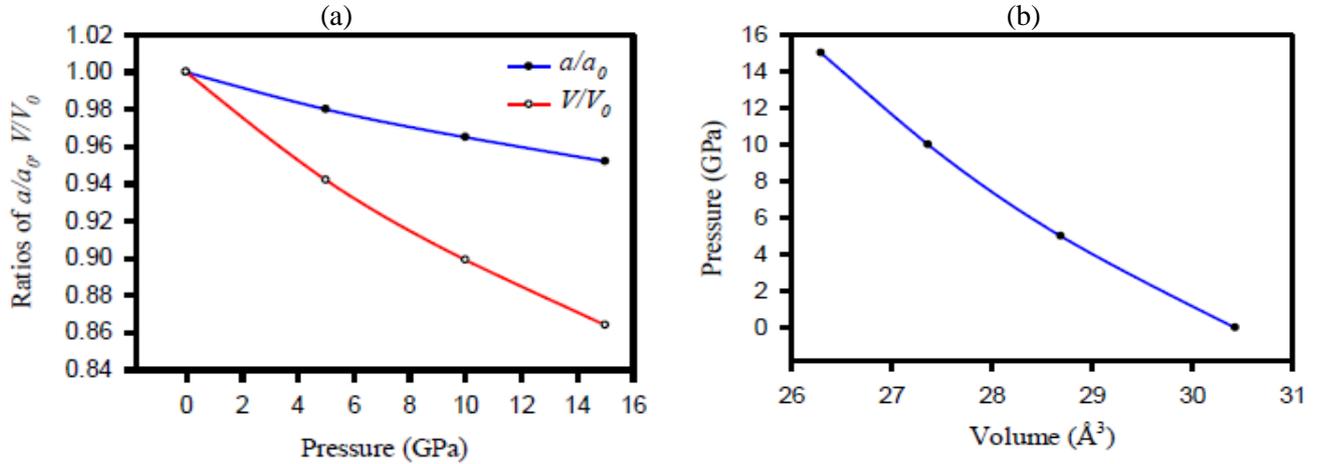

**Fig.2.** Variation of cell volume and lattice parameter as function of pressure (a), Birch-Murnaghan equation of state for MgCu (b).

## 3.2. Elastic Properties

Elastic constants are very crucial material parameters. The study of elastic constants provides a link between the mechanical properties and dynamic information concerning the nature of the forces operating in solids, especially for the stability and stiffness of materials [24]. It is very important to comprehend the Debye temperature, chemical bonds, and the mechanical stability of materials. The elastic constants relate to various fundamental solid-state phenomena such as stiffness, brittleness, stability, ductility, and anisotropy of material and propagation of elastic waves in normal mode. Hence it is meaningful to investigate the elastic constants in order to understand the mechanical properties of MgCu intermetallic compound at different pressure. The elastic constants are determined from a linear fit of the calculated stress-strain function according to Hook's law [25]. Since the intermetallic MgCu belongs to the cubic crystal, it has three independent elastic constants $C_{11}$, $C_{12}$ and $C_{44}$. In table 2, we have listed the calculated elastic constants of MgCu at pressure up to 15 GPa. It can be seen from Table 2 that the calculated results in this work are in good agreement with others theoretical values. Some variation appears due to the different calculation method.

**Table 2.** Calculated elastic constants $C_{ij}$ (GPa) and Cauchy pressure ($C_{12} - C_{44}$) of MgCu under hydrostatic pressure.

| $P$ (GPa) | $C_{11}$ | $C_{12}$ | $C_{44}$ | $C_{12} - C_{44}$ |
|---|---|---|---|---|
| 0 | 122.31 (119.47[a], 128.14[b]) | 54.61 (44.74[a], 61.06[b]) | 79.76 (72.26[a], 85.09[b]) | -25.15 (-27.52[a], -24.03[b]) |
| 5 | 129.75 | 76.29 | 91.04 | -14.75 |
| 10 | 146.21 | 114.54 | 99.44 | 15.10 |
| 15 | 148.66 | 144.74 | 106.26 | 38.48 |

[a]Ref. 10 (Using GGA); [b]Ref. 10 (Using LDA)

For mechanically stable crystals, the independent elastic constants should satisfy the well-known Born stability criteria [26]. The Born stability criteria for cubic structures are

$$C_{11} > 0, \ C_{44} > 0, \ C_{11} - C_{12} > 0 \ \text{and} \ C_{11} + 2C_{12} > 0$$

From table 2, it is clear that our calculated elastic constants satisfy the above stability criteria up to 15 GPa, ensuring their respective mechanical stability under pressure. We notice that the value of $C_{12}$ becomes larger than the value of $C_{11}$ at 16 GPa. It certainly violates the stability criteria for cubic crystal. Hence we predict that this compound becomes unstable under pressure more than 15 GPa. Moreover, a phase change can be occurred above 15 GPa pressure. To the best of our knowledge, there are no experimental and other theoretical data available in literature for the elastic constants of MgCu under pressure for comparison. So our study should motivate for the experimental confirmation of this result in future.

Fig. 3(a) shows the pressure dependence of elastic constants $C_{ij}$ and it is noticed that $C_{11}$, $C_{12}$ and $C_{44}$ increase with applied pressure. Generally a longitudinal strain produces a change in $C_{11}$, since $C_{11}$ represents the elasticity in length, whereas $C_{12}$ and $C_{44}$ are shear constant and represents the elasticity in shape [27]. A transverse strain could cause a change in shape, but no change in volume [28]. In this work we notice that $C_{12}$ is more sensitive to pressure than $C_{11}$ and $C_{44}$. This indicates that the shear constant is more sensitive than the young modulus.

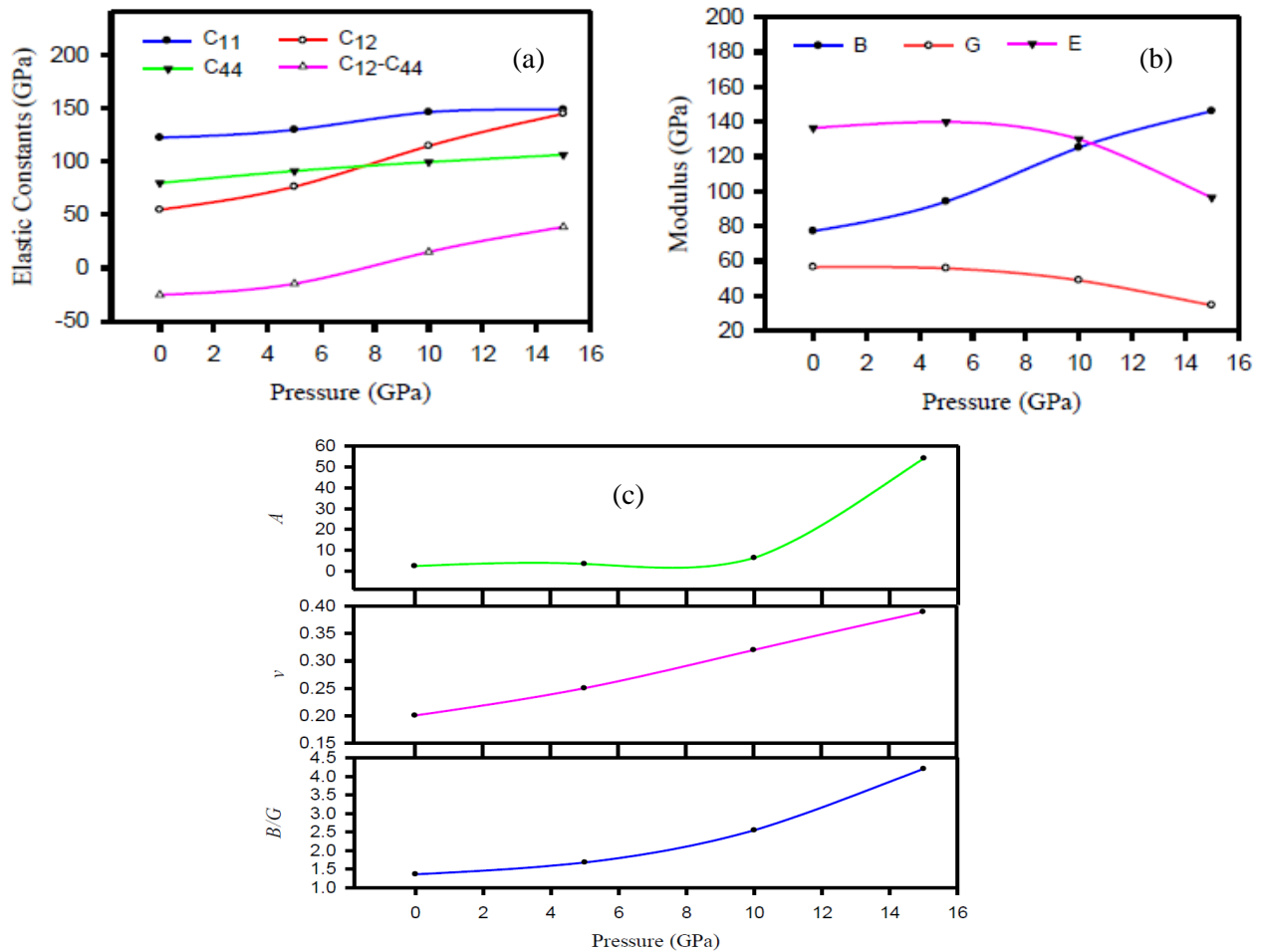

**Fig. 3.** The calculated elastic constants of MgCu under different pressure: (a) Elastic constants $C_{ij}$ and Cauchy pressure, (b) Elastic modulus, (c) *B/G*, Poisson's ratio and anisotropy factor.

$C_{12} - C_{44}$ is defined as Cauchy pressure which could be used to describe the angular character of atomic bonding in metals and compounds [29]. The negative value of Cauchy pressure indicates that the material is nonmetallic with directional bonding and if Cauchy pressure is positive, then the material is expected to be metallic [27]. From Table 2, it is evident that the value of the Cauchy pressure of MgCu at an ambient condition is negative indicates the nonmetallic behavior of this compound. But this result is totally opposite to the result getting from the band structure diagram. The explanation for this contradiction is currently unknown and further investigation is expected in future. However the Cauchy's pressure increases with the increasing of pressure and becomes positive at 10 GPa as shown in Fig. 3(a). Therefore, according to the above criterion it is noticeable that the metallic behavior of MgCu is getting stronger with increasing pressure. Furthermore if the Cauchy's pressure is negative, the material is expected to be brittle and if positive then the ductile nature of material exhibits [30]. In our present study this value is negative which indicates that MgCu is brittle at ambient pressure. Moreover, Cauchy pressure increases with the increasing of pressure and at pressure 10 GPa its value becomes positive as shown in Fig. 3(a). This result indicates that MgCu transformed into ductile nature from the brittle nature with the increasing of pressure.

Another index to explain the ductility and brittleness of a material is Pugh's ratio is defined as *B/G* [31]. A material behaves in a brittle manner, if *B/G* < 1.75, otherwise it should be ductile. From Table-3, it is evident that this compound is brittle at ambient pressure and with increasing pressure the ductile nature is exhibited gradually as shown in Fig. 3(c). Thus these same results of brittleness and ductility by investigating the Cauchy's pressure and Pugh's ratio ensure the reliability of our present study.

From results of the calculated $C_{ij}$, the most important parameters for estimating the mechanical properties of compounds such as the bulk modulus *B*, shear modulus *G*, Young's modulus *E* and Poisson's ratio *v* of MgCu are determined by using the Voigt-Reuss-Hill (VRH) averaging scheme [32] . They have crucial implication in engineering science. For the cubic system, the Voigt and Reuss bounds of *B* and *G* can be expressed as follows [33]:

$$B_v = B_R = \frac{(C_{11} + 2C_{12})}{3} \tag{1}$$

$$G_v = \frac{(C_{11} - C_{12} + 3C_{44})}{5} \tag{2}$$

$$G_R = \frac{5C_{44}(C_{11} - C_{12})}{[4C_{44} + 3(C_{11} - C_{12})]} \tag{3}$$

The Hill took an arithmetic mean value of *B* and *G* can be expressed as follows,

$$B = \frac{1}{2}(B_R + B_v) \tag{4}$$

$$G = \frac{1}{2}(G_v + G_R) \tag{5}$$

Young's modulus (*E*), and Poisson's ratio (*v*) can be calculated by using following relations,

$$E = \frac{9GB}{3B + G} \tag{6}$$

$$v = \frac{3B - 2G}{2(3B + G)} \tag{7}$$

The calculated values of *B, G, E, v*, and *B/G* at different pressures up to 15 GPa are tabulated in Table 3.

**Table 3.** The calculated bulk modulus *B* (GPa), shear modulus *G* (GPa), Young's modulus *E* (GPa), *B/G* values, Poisson's ratio *v* and anisotropy factor *A* of MgCu compound under hydrostatic pressure.

| *P (GPa)* | *B* | *G* | *E* | *B/G* | *v* | *A* |
|---|---|---|---|---|---|---|
| 0 | 77.17 | 56.54 | 136.32 | 1.36 | 0.20 | 2.35 |
| 5 | 94.11 | 55.85 | 139.87 | 1.68 | 0.25 | 3.40 |
| 10 | 125.10 | 48.97 | 129.95 | 2.55 | 0.32 | 6.27 |
| 15 | 146.04 | 34.65 | 96.33 | 4.21 | 0.39 | 54.21 |

From Table 3, it can be seen that $B > G$, which indicates that the shear modulus is the prominent parameter associating with the stability of cubic MgCu [34]. The calculated bulk modulus *B* at pressure 0 GPa agrees well with the Bulk modulus $B_0$, listed in Table 1, which was obtained through the fit to a Birch-Murnaghan EOS. Some variation appears due to different calculation method. This means that the approach of the calculation is reliable. The bulk modulus is usually considered to be a measure of resist deformation capacity upon applied pressure. The larger the value of bulk modulus is, the stronger capacity of the resist deformation is. From Fig. 3(b), we can see that the value of *B* increases with the increase of the pressure, meaning that the existence of external pressure increases the capacity of the resist deformation of MgCu. On the other hand, the shear moduli and Young's modulus are the measure of resist reversible deformation by shear stress and stiffness of the solid materials respectively. From Fig. 3(b), one can see that the value of *G* and *E* decrease with the increase of external pressure. This means that, the capacity of the resist shear deformation and stiffness of MgCu are decreased with the increase of pressure respectively. The Poisson's ratio is used to reflect the stability of the material against shear and provides information about the nature of the bonding forces [35]. Bigger the Poisson's ratio betters the plasticity. The value of *v* for covalent materials is small ($v = 0.1$), and for ionic materials 0.25. The value between 0.25 and 0.5 indicates that the force exists in the solid is central [36]. From Fig 3(c), we see that, at ambient condition the value of *v* is 0.20, which is very near to the value 0.25 and at pressure 5 GPa the value becomes exactly 0.25, indicating ionic nature is dominant of MgCu. Above pressure 5 GPa to 15 GPa, the values are in the range which satisfy the condition to be the interatomic forces central of MgCu compound. The Zener anisotropy factor *A* is the measure of the degree of anisotropy in solid [37]. $A = 1$ means a completely isotropic material, whereas a value smaller or larger than unity indicates the degree of elastic anisotropy. The value of *A* of MgCu is calculated by using the following formula at different pressures and listed in Table 3.

$$A = \frac{2C_{44}}{(C_{11} - C_{12})} \tag{8}$$

As shown in Table 3, the value of A for MgCu is larger than unity, indicating that the material under studied can be regarded as elastically anisotropic material. From Fig. 3(c), it can be seen that the value of *A* increases with the increase of pressure up to 15 GPa. On other words, the degree of elastic anisotropy is getting larger with the increase of pressure of compound MgCu.

### 3.3. Electronic properties

The density of states (DOS) plays vital role in the analysis of the physical properties of materials. In order to get a further insight into the bonding characteristics of MgCu, the total density of states (TDOS) and partial density of states (PDOS) of MgCu have been calculated. The electronic structure reveals the mechanism about structural stability and elastic properties of MgCu compound. The calculated electronic band structures of MgCu along the high symmetry directions in the Brillouin zones are shown in Fig. 4. Fig. 5(a) depicts the TDOS and PDOS of MgCu at zero pressure in the energy range from -45 eV to 25 eV, while Fig. 5(b) illustrates the TDOS as a function of pressure.

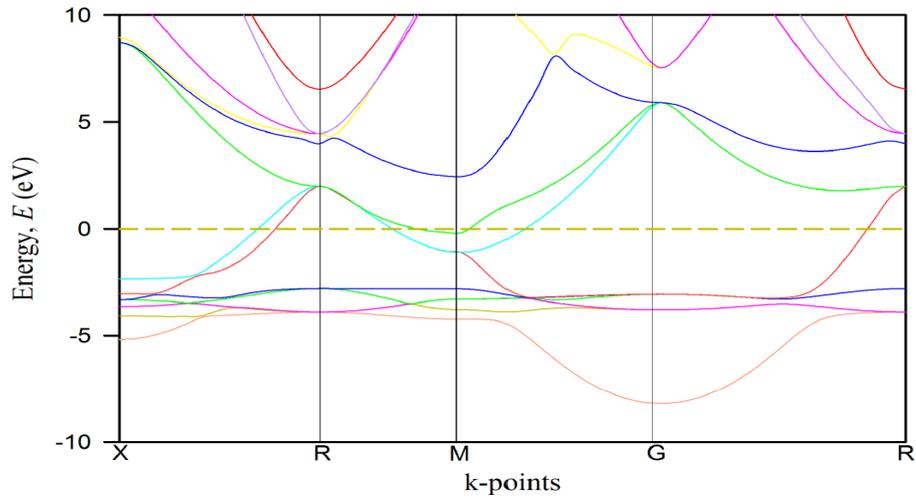

**Fig. 4.** Electronic band structure of MgCu along high symmetry direction in the Brillouin zones.

From Fig. 4, it is noticed that the compound under study is metallic because a number of valance and conduction bands are overlapping at the Fermi level. From Fig. 5(a), it is observed that the main bonding peaks locates in the energy range between Fermi level and -9 eV, in which the most dominant contribution comes from the Mg-p and Cu-d states. But at Fermi level the contribution of Cu-d orbital is dominant. In case of Cu, the DOS profile is relatively low and flat at higher energy level due to fully occupied d orbits of Cu. Structural stability of intermetallic compound is associated with its bonding electronic orbits [38]. For ionic bond, it depends on the charge transfer between atoms, while covalent bond is determined by the depth and width of band gap near Fermi level. Based on this information, the formation of covalent bond in MgCu compound is very weak. For further understanding about the bonding property of MgCu, we have run further investigations on Mulliken overlap population [39]. Mulliken overlap population is a great quantitative criterion for investigating the covalent and ionic nature of bonds. In Table 4, we have listed the atomic Mulliken population of MgCu compound. A high value of the bond population indicates a covalent bond, whereas a low value denotes the ionic bonds. A value of zero indicates a perfectly ionic bond and the values greater than zero indicate the increasing levels of covalency [40].

**Table 4.** Mulliken electronic populations of MgCu.

| Species | s | p | d | Total | Charge | Bond | Population | Lengths |
|---|---|---|---|---|---|---|---|---|
| Mg | 0.35 | 6.45 | 0.00 | 6.61 | 1.19 | Mg-Cu | -0.76 | 2.780 |
| Cu | 0.94 | 1.49 | 9.76 | 12.19 | -1.19 | | | |

From Table 4, it is found that for MgCu, the charge transfer from Mg atoms to Cu atoms. The bond population is -0.76 exhibits the ionic nature of MgCu compound which is agreeable with the result having form the calculation of elastic constant.

In order to understand the variation of the TDOS of MgCu with applied pressure, the total density of states (TDOS) of MgCu at pressure 0, 5 and 15 GPa is also investigated, as displayed in Fig. 5(b). Investigating Fig. 5(b), we don't have any changes of the shapes of the peaks. Hence we further investigate the TDOS within short energy range as shown in Fig. 5(c). Analyzing Fig. 5(c) it is found that the shapes of the peaks changes slightly under pressure, which means the structure of MgCu compound has no dramatic changes and there is no structural phase transformation under the pressure up to 15 GPa. We observe that with increasing pressure the peaks are shifted toward the Fermi level.

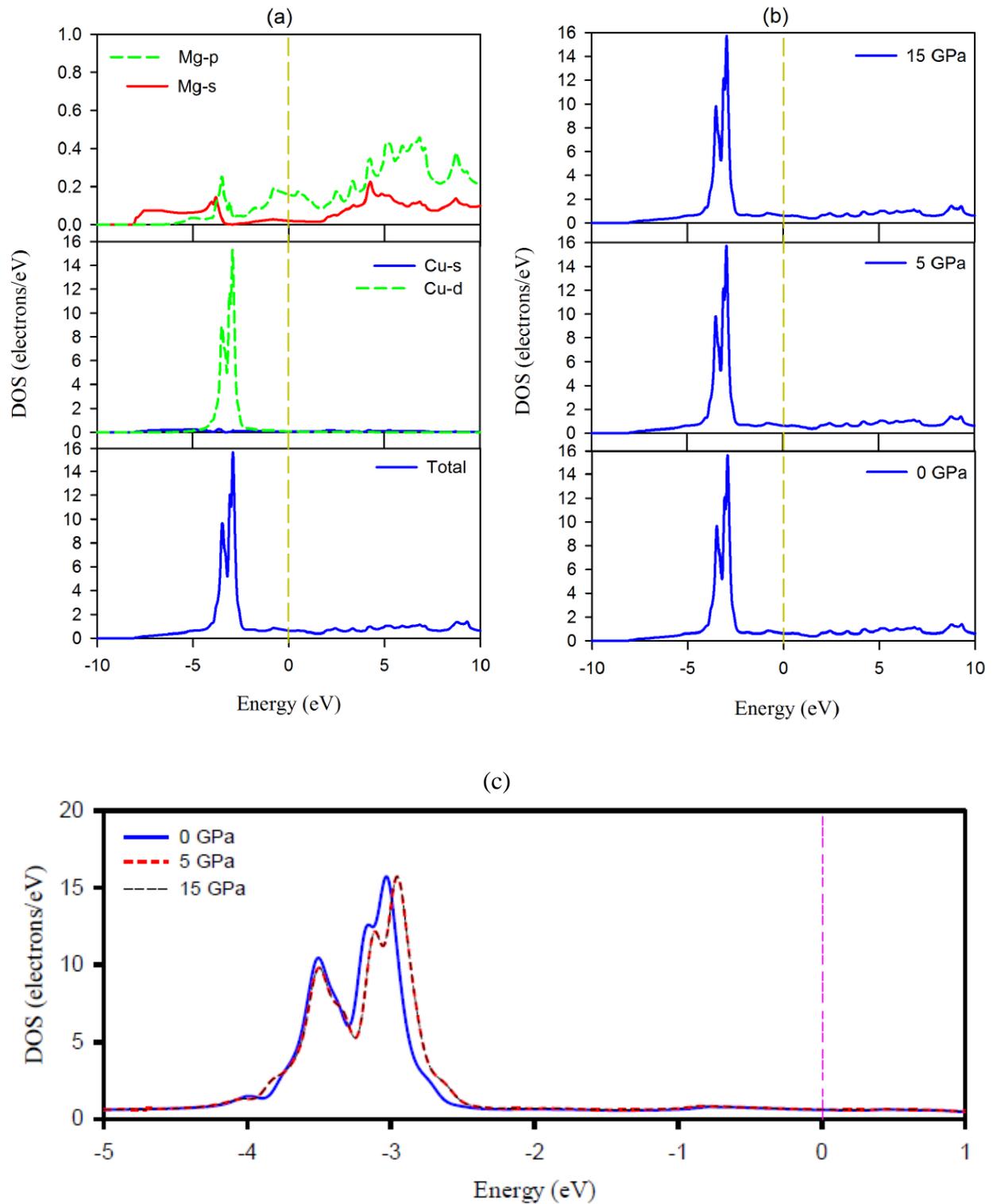

**Fig. 5.** The density of states of MgCu under different pressure (a) zero pressure, (b) high pressure and (c) high pressure (short energy range).

The decreases or increases of DOS at Fermi level are corresponded to the hardening or softening of the materials respectively [41]. The calculated DOS at 0, 5 and 15 GPa was 0.59, 0.61 and 0.61 electrons/eV respectively, which indicates the softness of MgCu compound has been improved by increasing pressure. This result is consistent with the analysis of the elastic properties of MgCu. Overall it can be concluded that both ionic and covalent bond exists in MgCu where ionic nature is dominant of MgCu and with increasing pressure both bonds become weak and the material become soft gradually.

## 3.4. Optical properties

The optical properties of MgCu is calculated by the frequency-dependent dielectric function, $\varepsilon(\omega) = \varepsilon_1(\omega) + i\varepsilon_2(\omega)$. The imaginary part $\varepsilon_2(\omega)$ is obtained from the momentum matrix elements between the occupied and the unoccupied electronic states and calculated directly using [16],

$$\varepsilon_2(\omega) = \frac{2e^2\pi}{\Omega\varepsilon_0} \sum_{k,v,c} |\psi_k^c| u.r |\psi_k^v|^2 \delta(E_k^c - E_k^v - E) \tag{9}$$

Where, $u$ is the vector defining the polarization of the incident electric field, $\omega$ is the frequency of light, $e$ is the electronic charge and $\psi_k^c$ and $\psi_k^v$ are the conduction and valence band wave functions at $k$, respectively. The real part is derived by using the Kramers-Kronig transform. All other optical constants, such as refractive index, loss-function, absorption spectrum, reflectivity and conductivity are those given by Eqs. 49 to 54 in Ref. [16]. Fig.6 exhibits the optical functions of MgCu calculated for photon energies up to 80 eV for polarization vector [100]. We have used a 0.5 eV Gaussain smearing for all calculations.

Fig. 6(a) represents the reflectivity spectra of MgCu as a function of photon energy. We notice that the reflectivity is 0.53-0.67 in the infrared region (1.24 meV-1.7 eV) and the value drops rapidly in the visible region and again increases in ultraviolet region up to 72 eV (reaching maximum at 72 eV). The compound with high reflectivity in the high energy region can be used as a good coating material to avoid solar heating.

The absorption coefficient provides data about optimum solar energy conversion efficiency and it indicates how far light of a specific energy (wavelength) can penetrate into the material before being absorbed [42]. Fig. 6(b) presents the absorption coefficient spectra of MgCu. The absorption spectra exhibit several peaks but the hight peak located at 64.27 eV indicates rather good absorption coefficient in the ultraviolet region for MgCu.

For designing photoelectric device the complex refractive index is crucial. Fig. 6(c) illustrates the refractive index of MgCu as a function of photon energy. The static refractive index of MgCu is 6.55. From the Fig. 6(c) it is clear that the refractive index is high in the infrared region and gradually decreased in the visible and ultraviolet region.

Fig 6(d) illustrates the dielectric function of MgCu as a function of photon energy. It is observed from Fig. 6(d) that the value of $\varepsilon_2$ becomes zero at about 72 eV indicating that the material becomes transparent above 72 eV. Generally the value of $\varepsilon_2$ becomes nonzero when absorption starts. Hence it is clear from Fig. 6(d) that absorption occurs between 0-19.87 eV, and 41.92-67.05 eV, which is also evident from Fig. 6(b). The value of the static dielectric constant is 43, indicating that MgCu is a promising dielectric material.

Fig. 6(e) shows the conductivity spectra of MgCu at normal pressure and exhibits the metallic nature of MgCu, since the photoconductivity begins with zero photon energy. This result ensures the validity of our current DFT based calculations. However, photoconductivity (electrical conductivity) of MgCu increases due to the result of absorbing photons [43].

Fig. 6(f) represents the energy loss spectrum as a function of photon energy, which is a crucial optical parameter to explain the energy loss of a fast electron traversing the material and is large at the plasma frequency [44]. We observed several peak but the prominent peak found at 6 eV, which corresponds to the rapid demission in the reflectance.

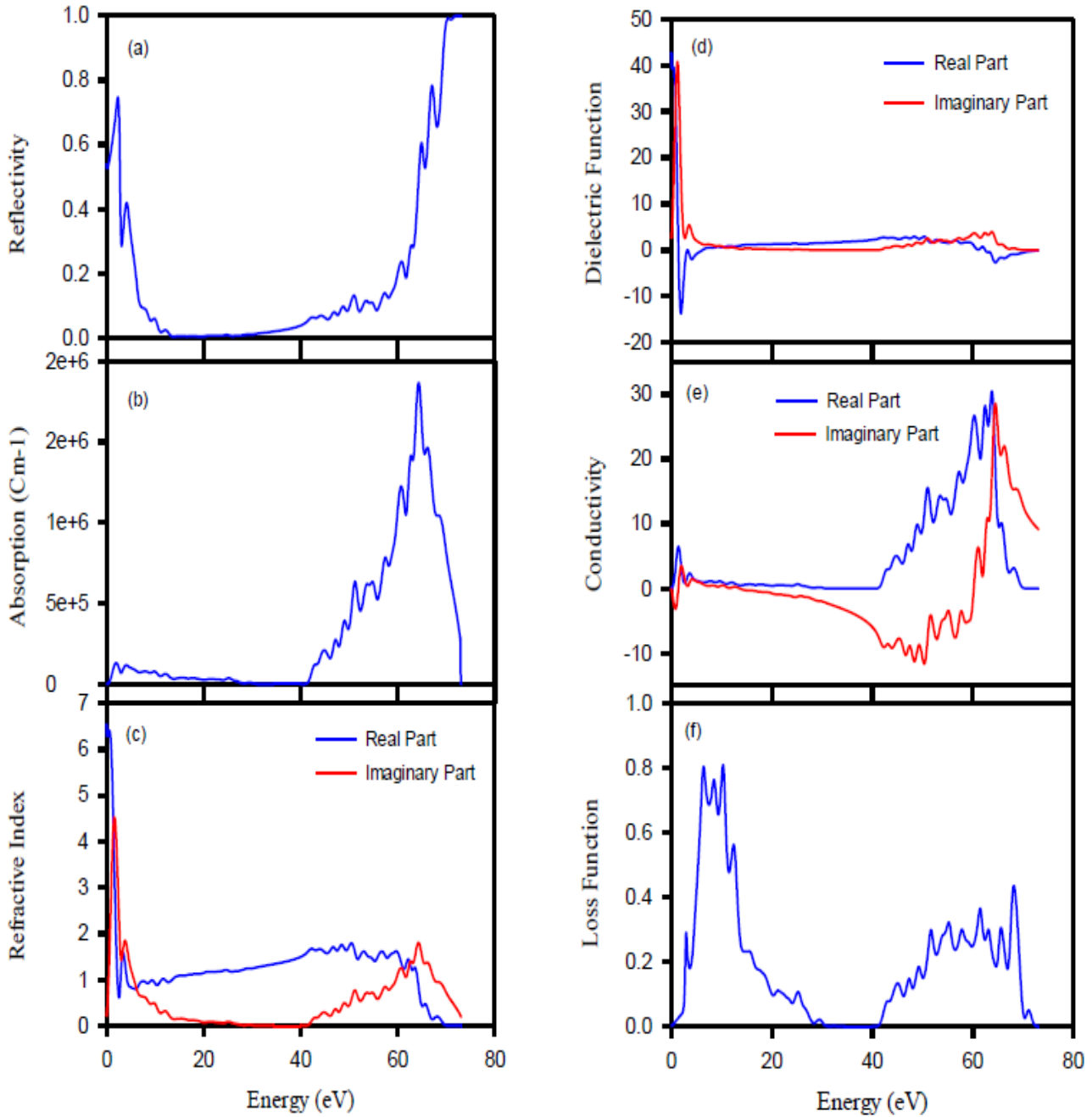

**Fig. 6.** The optical functions (a) reflectivity, (b) absorption, (c) refractive index, (d) dielectric function, (e) conductivity, and (f) loss function of MgCu for polarization vector [100].

# 4. Conclusions

In this paper, we have investigated the structural, elastic, electronic and optical properties of intermetallic compound MgCu under various pressures up to 15 GPa by performing the generalized gradient approximation (GGA) in the frame of density functional theory. The main conclusions are as follows:

(1) The obtained lattice parameters are in good agreement with the experimental and previous theoretical values. It may be worthwhile to note that the normalized structural parameters $a/a_0$ and $V/V_0$ decreases with the increase of pressure.
(2) The study of elastic properties shows that the structure of MgCu intermetallic compound is mechanically stable with the pressure range from 0 to 15 GPa. It is worthy to mention that the material becomes unstable under the pressure more than 15 GPa. The study of bulk modulus $B$, shear modulus $G$ and Young's modulus $E$ under various pressures show that $B$ increases with the increase of the pressure, meaning that the existence of external pressure increases the capacity of the resist deformation of MgCu whereas $G$ and $E$ decrease with the increase of external pressure. The study of Cauchy pressure and Pugh's ratio exhibit that the compound is transformed into ductile nature from brittle nature with the increase of pressure.
(3) The electronic band structure shows that the MgCu compound exhibits metallic characteristics. The result of DOS shows that there is no structural phase transformation under pressure ranges from 0 to 15 GPa. Further study on the bonding nature of MgCu shows that both ionic and covalent bond exists in MgCu where ionic nature is dominant. The calculated DOS at Fermi level under different pressure indicates that the softness of MgCu compound can be improved by increasing pressure.
(4) The reflectivity spectrum shows that the reflectivity is high in the ultraviolet region up to 72 eV. This compound shows rather good absorption coefficient in the ultraviolet region. The conductivity spectrum shows that electrical conductivity of this material increases as a result of absorbing photons.